# Nanostructure engineering of epitaxial piezoelectric α-quartz thin films on silicon

Q. Zhang[1, 2 γ], D. Sánchez[1 γ], R. Desgarceaux[1 γ], A. Gomez[2], P. Escofet-Majoral[1], J. Oró-soler[2], J. Gazquez[2], G. Larrieu[3], B. Charlot[1], M. Gich[2], A. Carretero-Genevrier[1]*

1. Rudy Desgarceaux, Pau Escofet-Majoral, Dr. Benoit Charlot, Dr. Adrian Carretero-Genevrier
Institut d'Electronique et des Systemes (IES), CNRS, Université de Montpellier, 860 Rue de Saint Priest 34095 Montpellier, France
2. Qianzhe Zhang, Andres Gomez, Dr. Jaume Gazquez, Dr. Marti Gich
Institut de Ciència de Materials de Barcelona ICMAB, Consejo Superior de Investigaciones Científicas CSIC, Campus UAB 08193 Bellaterra, Catalonia, Spain
3. Guilliem Larrieu
LAAS-CNRS, Université de Toulouse, CNRS, INP, 7 av. Colonel Roche, 31031Toulouse, France

γ These authors contributed equally to this work
E-mail: carretero@ies.univ-montp2.fr,

**Abstract**

The monolithic integration of sub-micron quartz structures on silicon substrates is a key issue for the future development of telecommunication to the GHz frequencies. Here we report unprecedented large-scale fabrication of ordered arrays of piezoelectric epitaxial quartz nanostructures on silicon substrates by the combination of soft-chemistry and three cost effective lithographic techniques: (i) laser transfer lithography, (ii) soft nanoimprint lithography on Sr-doped $SiO_2$ sol-gel thin films and (iii) self-assembled $SrCO_3$ nanoparticles reactive nanomasks. Epitaxial α-quartz nanopillars with different diameters (down to 50 nm) and heights (up to 2000 nm) were obtained for the first time. This work proves the control over the shape, micro- and nano-patterning of quartz thin films while preserving its crystallinity, texture and piezoelectricity. This work opens up the opportunity to fabricate new high frequency resonators and high sensitivity sensors relevant in different fields of application.

1. **Introduction.**

Piezoelectric materials are elements of motion sensors (accelerometers and gyroscopes), oscillators and resonators present in practically any single electronic circuit. As a result, the piezoelectric materials solves many technological challenges including high frequency and stable oscillators, reducing energy consumption of devices and incorporating simple and efficient inertial sensors for distance, movement and acceleration detection. The integration of high quality epitaxial piezoelectric films and nanostructures on silicon is a milestone towards the expansion of novel devices with the traditional Si-based complementary metal-oxide-semiconductor (CMOS) technology[1]. In addition, advances in micro and nanofabrication technologies, open the possibility to implement a large scale integration of miniaturized piezoelectric materials into innovative electromechanical devices with nanosized moving parts with prospective sensor applications in electronics, biology and medicine[2].

In this context, α-quartz is widely used for electronic applications: its piezoelectric properties allow for an excellent control of the frequency in oscillators and for producing very selective filters[3]. Since the eigen frequency of the quartz crystal is very sensitive to changes of its mass or acceleration, this material is extremely convenient to implemented micro-resonators for sensing applications (strength, humidity, acceleration, etc.) [4]. However, α-quartz and other piezoelectric sensing materials having extremely large quality factor ($Q > 10^6$)[5], high temperature stability and very low phase noise are only available as bulk single crystals. For this reason, all these excellent sensing materials can only be configured as high performance transducers through direct bulk micromachining or hybrid integration methods[6]

Nowadays, the lowest achievable thickness of quartz crystals is about 10 μm and 100 μm in diameter, which in turn limits the working frequencies of the transducers (the resonance frequency corresponds to half of the wavelength of the crystal thickness). few works have shown sub-micron patterned quartz surfaces, such as those prepared by Laser Interference Lithography[7] and by Faraday cage angled-etching technique[8] or lithium niobiate nanostructures synthetized by focused ion beam (FIB) technology[9], although from bulk single crystals.

We recently developed the direct and *bottom-up* integration of epitaxial α-quartz thin films on silicon substrate by chemical solution deposition (CSD)[10], which overcomes the aforementioned limitations. The method relies on the thermal devitrification and crystallization of dip-coated mesoporous silica films, assisted by strontium alkaline earth cation in amphiphilic molecular templates[10a]. This new approach opens the door for developing efficient quartz-based piezoelectrics devices engineered from widely available and non-toxic compounds using industrially scalable methods.

In the present work, we have taken advantage of an improved evolution of this chemical route[10a] and we have combined it with a set of cost-effective *top-down* lithography techniques to fabricate large scale epitaxial nanopatterned quartz thin films on silicon substrates with controllable nano and microstructures. This work is, to our knowledge, the first example that shows the possibility of engineering the integration on silicon of patterned quartz thin films, which precedes the production of nanostructured and microelectromechanical systems, as previously highlighted[11]. By engineering nanostructured quartz films on silicon, which are much thinner (200 nm - 1 µm) than those obtained by top down technologies on bulk crystals, one can expect higher resonance frequencies. In addition, the control of the porosity, texture, shape, micro- and nano-patterning of quartz thin films opens up the opportunity to produce more efficient devices. This is supported by the fact that nanostructured quartz thin films increase the specific area thus, enhancing the sensing properties of the future device.

2. **Results and discussions**

To produce nanoscaled 1D (arrays of pillars) on epitaxial α-quartz thin films by silicon micromachining, we have tested cost efficient lithographic techniques such as laser transfer lithography technique[12], Soft nanoimprint lithography[13] and a novel plasma-assisted self-assembled SrCO3 nanoparticles reactive nanomask etching. Such procedures do not require any lithographic mask and allow obtaining a large scale and precise control of epitaxial quartz nanostructures (see Fig. 1).

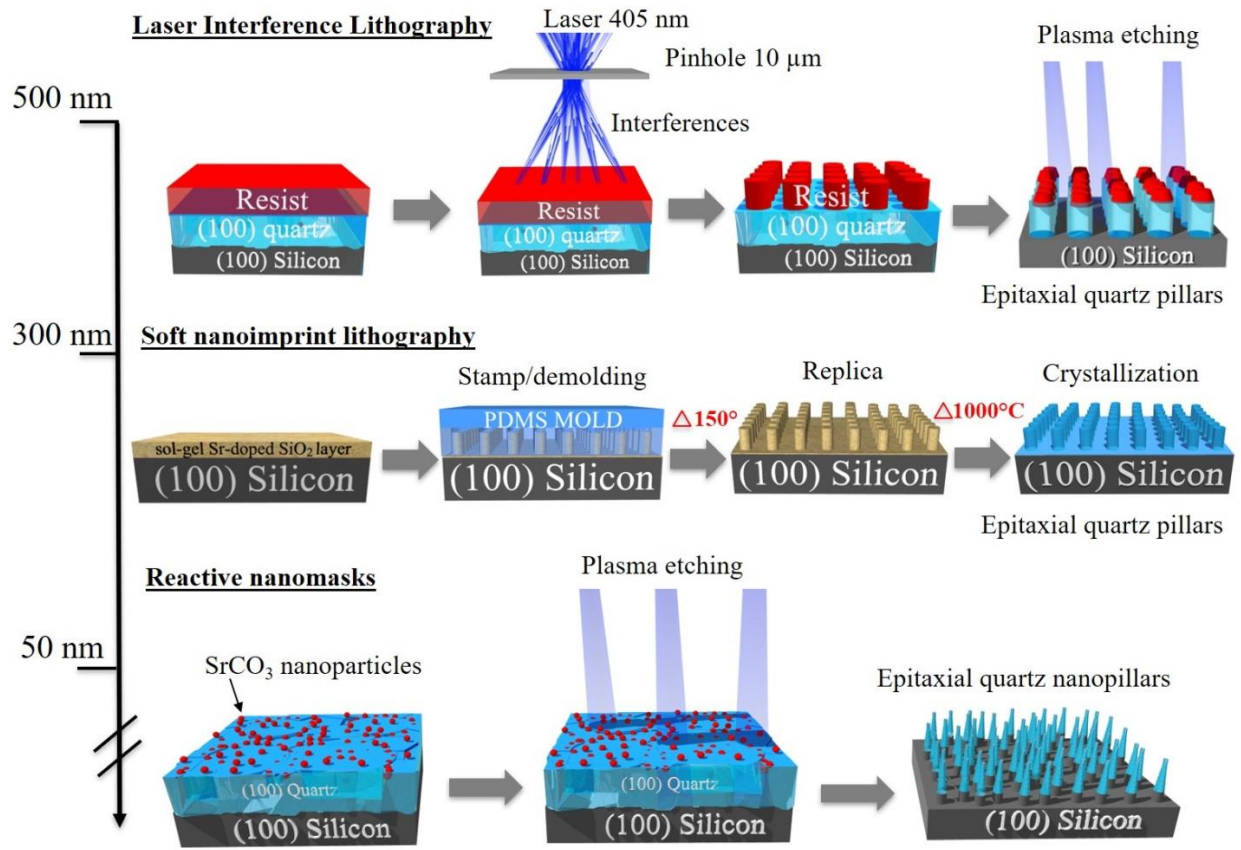

**Fig. 1. Schematics that summarizes the key steps that were applied to produce nano lithographic patterns on epitaxial quartz thin films dependent of the film thickness.**

We initially produced micrometric photolithographic patterns on 300 nm thick epitaxial quartz films on silicon (100) to evaluate the crystal stability of quartz under an anisotropic plasma etching process (see Fig.SI1). In this case, 5 µm wide lines of epitaxial quartz were achieved, which did not undergo amorphization during the lithographic process. This feature was confirmed by 2D X-ray diffraction (XRD) and Piezo-response Force Microscopy (PFM) on micro-photolithographed samples (see Fig. SI1). The XRD analysis revealed the same (100) α-quartz out of plane texture that the films had before the etching process (see Fig. SI1). PFM measurements showed that the piezoelectricity of micro patterned quartz films was preserved (see Fig.SI1 d and e), with the piezoelectric coefficient ($d_{33}$) being comparable to that of the quartz bulk material (i.e. 1.5 and 3.5 pm/V)[14].

### 2.1. Direct patterning of epitaxial piezoelectric quartz thin films by Laser Interference Lithography.

Next, we used Laser Interference Lithography (LIL, also known as holographic lithography) to obtain direct sub-micro patterned epitaxial quartz films. LIL process is a top-down fabrication technique that is currently used to selectively pattern single crystals into dense vertical nanocolumn arrays[7]. This technique allows generating arrays of lines or dots in a photoresist film from an interference pattern generated by a UV laser over $cm^2$ surfaces and with pitches ranging between 400 nm up to 2400 nm (see Fig. 1). The mask-less exposure of the photoresist layer together with two or more coherent light beams offers a simple, low-cost, large-area nanolithography technique[12].

Figure 2a shows a first nano lithographic pattern on epitaxial quartz thin films on (100) silicon substrate using LIL lithography. Low resolution Scanning Electron Microscopy image (SEM) displays an ultra-dense network of quartz nanocolumns with a precise control over their diameter, height and position. The Transmission Electron Microscopy image of the α-quartz/Si interface in a 600 nm height nanopillar and the its corresponding Electron Diffraction pattern confirm that the crystalline quality of epitaxial α-quartz has been preserved during the LIL and RIE etching process (see Fig.2b).

Piezoelectric coefficient $d_{33}$ estimated by PFM on quartz nanocolumns were compared with those values obtained in dense quartz films before the lithographic process using an alternative method, direct piezoelectric force microscopy (DPFM), recently developed by the A. Gomez *et al.*[15] (see Fig. SI2 and Fig. 2c). We employed DPFM measurements to obtain the piezoelectric characterization of dense quartz film and compared the results with the case of nanocolumns. The piezoelectric properties of nanopatterned epitaxial quartz films were measured by using PFM technique in 800 nm thick nanocolumns (see Fig. 2c). Both piezoelectric coefficient ($d_{33}$) are similar and comparable to that of the quartz bulk material[14] (i.e. $d_{33(PFM)} = 2\pm0.5$ pm/V and $d_{33(DPFM)} = 4\pm2$ pC/N) confirming that the piezoelectric functionality of nanocolumns is preserved (see Fig. 2d and S2, respectively). Notice that all DPFM measurements were compared with a reference based in a commercial ferroelectric Periodically Poled Lithium Niobate (PPLN) sample (see for more details Fig. SI3).

Graphical representation in Figure 2d shows the control of quartz nanocolumns height (up to 800 nm) as a function of the number of deposition of $SiO_2$ layers. The followed multilayer film synthesis consists on the sequential deposition and consolidation of several gel layers before a final annealing treatment aiming to induce the epitaxial α-quartz growth (see more details in experimental section). Fig. S4a presents a series of FEGSEM images corresponding to the cross sections of dense quartz films before the lithographic process consisting of 1 up to 5 layers. Notice that the thicknesses of the different multilayers films correspond to the final quartz columnar heights plot in figure 2d. From the XRD measurements of Fig. S4b we can see that all the films present the usual α-quartz (100) out of plane texture and thus, the intensity of the reflections is proportional to the number of layers of the film. This demonstrates that the multi-layer approach allows controlling the nanocolumn height produced by LIL lithography while maintaining the crystallinity and crystal orientation. Moreover, the long range θ-2θ XRD pattern and pole figure of a 5-layer lithographed film confirms the texture of the (100) α-quartz crystallographic phase after the lithographic process. Nanocolumns conserve the (100) α-quartz∥(100) Si epitaxial relationship previously observed in dense films[10a]. Besides, no supplementary peaks from other reflections or polycrystallinity signals appear in the θ-2θ.

With the combination of our multi-layer deposition approach with LIL lithography, we have produced high aspect ratio epitaxial quartz columns with micrometric heights from dip-coated films. This was possible because the multi-layer deposition approach circumvented the maximum achievable thickness imposed by lateral tensile stresses that appeared during the densification of the layers[16]. A key step to overcome this obstacle has been to perform a temperature treatment to consolidate the gel layer (450ºC for 10 min in air atmosphere) after each deposition (See more details in the experimental section).

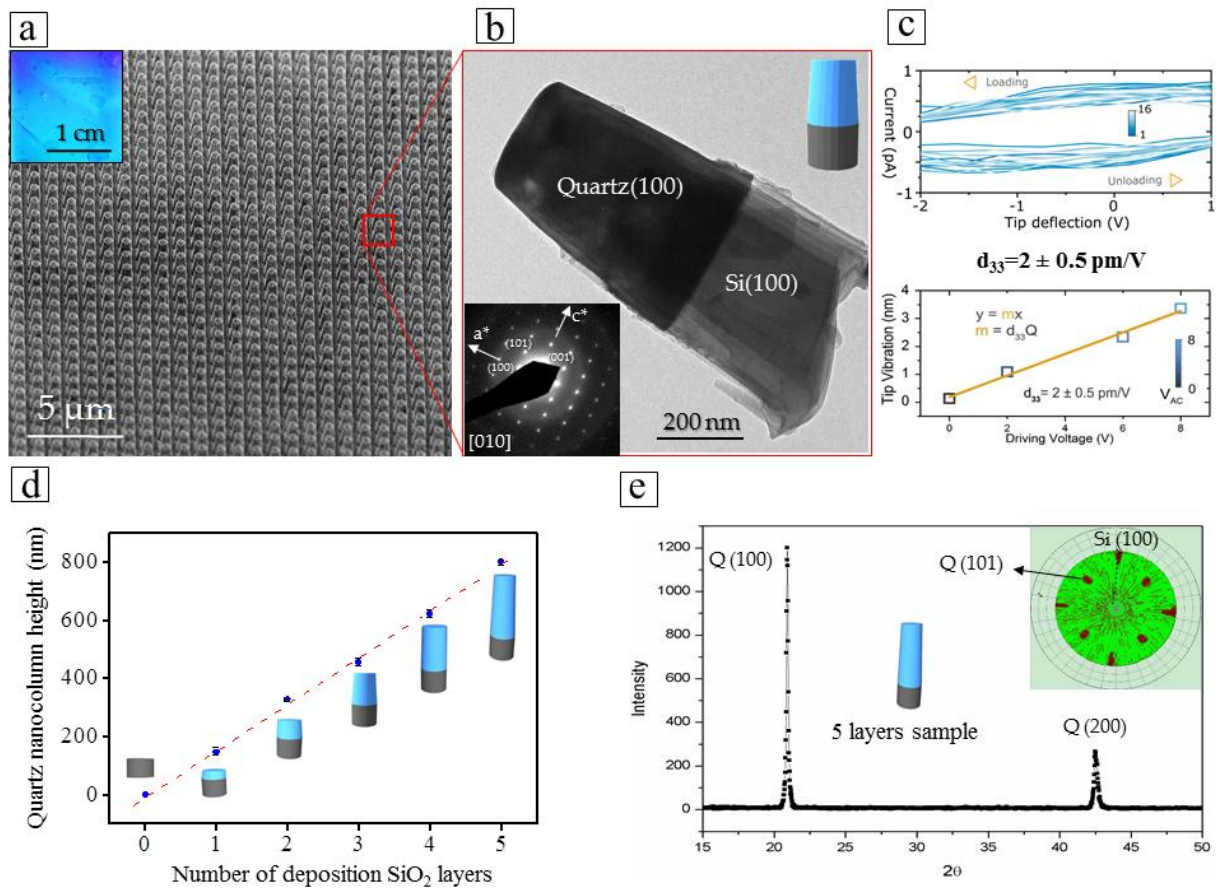

Fig. 2. **1D Lithographic patterning of epitaxial quartz thin films using LIL process. (a)** SEM image of ultra-dense network of 600 nm thick epitaxial quartz nanocolumns. The inset image shows a low resolution optical image of the sample. **(b)** Transmission Electron Microscopy image and electron diffraction measurement (inset image) of the α-quartz/Si interface of a single nanocolumn. **(c)** DPFM spectroscopic measurements on a 800 nm height quartz film obtaining with a loading rate of 95000 µN/s. **(d)** Graphic that shows the control of quartz nanocolumns height with the number of multideposited silica solgel layers before crystallization. **(e)** Long range θ-2θ XRD pattern with a perfect texture of the (100) α-quartz crystallographic phase after lithographic process. The inset image shows a pole figure of a 5-layer lithographed quartz film with (100) α-quartz∥(100) Si epitaxial relationship.

## 2.2. Soft nanoimprint lithography on Sr-doped SiO₂ sol-gel to nanostructurate epitaxial quartz films.

As an alternative route to LIL lithographic process, we applied soft Nano-Imprint Lithography (NIL), which combines top-down and bottom-up (sol-gel) approaches in order to produce epitaxial quartz nanopillar arrays with a precise control of pillar diameters and heights and inter-pillar distances on silicon. We want to emphasize that with this methodology we have reached unprecedented heights of 2 µm (see Figs. 3a and S5). The experimental procedure consisted in the combination of dip-coating process to synthetize Sr-doped xerogel silica films of controlled thicknesses on (100) silicon substrates with LIL and Nano-Imprint (NIL) lithographic techniques. In a first top-down fabrication step, large scale Si (100) masters made of nanopillars arrays were obtained by using LIL lithography and transferred by reactive ion etching at low pressure. Then, a second step involved the preparation of high quality PolyDiMethylSiloxane (PDMS) molds from Si(100) masters (see Fig. 3a) that produce perfectly imprinted Sr-doped silica nano-pillars with controlled diameter and height on silicon, as illustrated in figure 3b (See more details in experimental section).

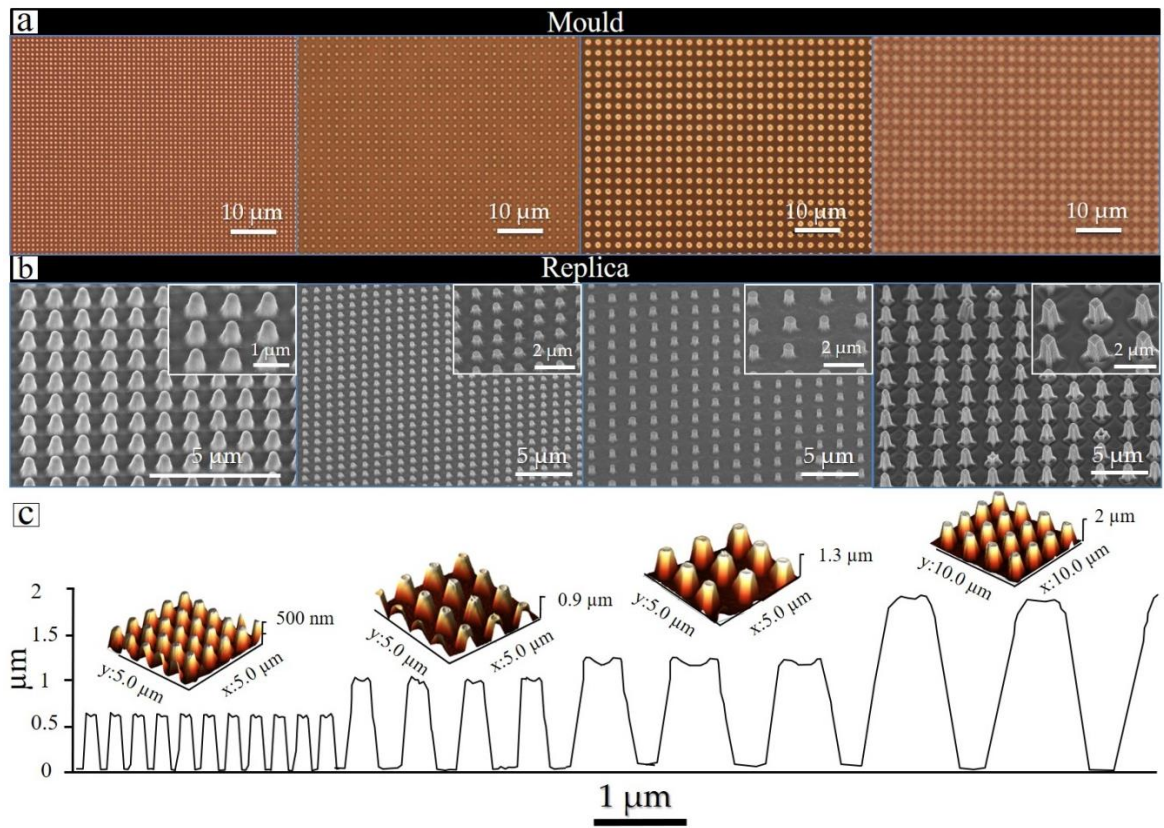

**Fig. 3. Optical images of Si (100) masters used along this work obtained by using LIL lithography (a)** FEG-SEM image of printed Sr-doped silica nano-pillars with controlled diameter (down to 350 nm) and height (up to 2 µm) on silicon. The inset pictures shows FEG-SEM images at higher magnification of pillars. **(b)** 3D AFM images showing silica nanostructured films prepared by NIL lithography in b. Below you can distinguish the profile analysis of the AFM image in c, revealing a perfect transfer of the different motives.

Finally, imprinted epitaxial (100) α-quartz nano-pillars arrays on silicon were obtained applying a thermal treatment at 1000ºC for 5 hours (see Fig. 4). Both optical microscope and SEM images shows Sr-doped silica xerogel nanopattern composed of 600 nm height columns before crystallization process, see Fig. 4a. The crystallized sample is shown in Fig. 4b, which exhibits the characteristic quartz grain boundaries at the nanostructured surface film (see optical image in figure 4b). Atomic resolution HAADF image of a single quartz nanocolumn/silicon interface reveals the epitaxial growth of quartz layer with an atomically sharp interface with the silicon substrate as shown in Fig. 4d (see also Figs. S6 in S.I.). In order to attain both a continuous nanostructured crystalline quartz film and a perfect nano-imprinted pattern, the dip coater deposition conditions have to be optimized. Likewise, a first mesoporous silica xerogel adhesion layer was needed to obtain an optimal print of PDMS molds on the Si(100) substrate (see Fig. 4c). This adhesion layer is consolidated at 450 °C during 5 min, before the deposition of the final printable silica layer. Both layers have the same thickness (200 nm) and are deposited under the same conditions i.e. at 25°C, 45% of humidity after applying a withdrawal speed of 300 mm min$^{-1}$ with the dip-coater. It is worth noting that the withdrawal speed of 300 mm min$^{-1}$ determines the thickness of the film[17] which, is indeed a critical factor to produce continuous nanostructured quartz layers. Below this critical withdrawal speed, the nucleation and crystallization of quartz layers was partial. As a result, fully crystallized films with a 100% surface coverage cannot be obtained for film thicknesses below 200 nm (see Fig. S7a). This trend is also reflected by the XRD patterns of the films which display higher intensities of the α-quartz (00L) reflections for increasing withdrawal speeds (see Fig. S7b)

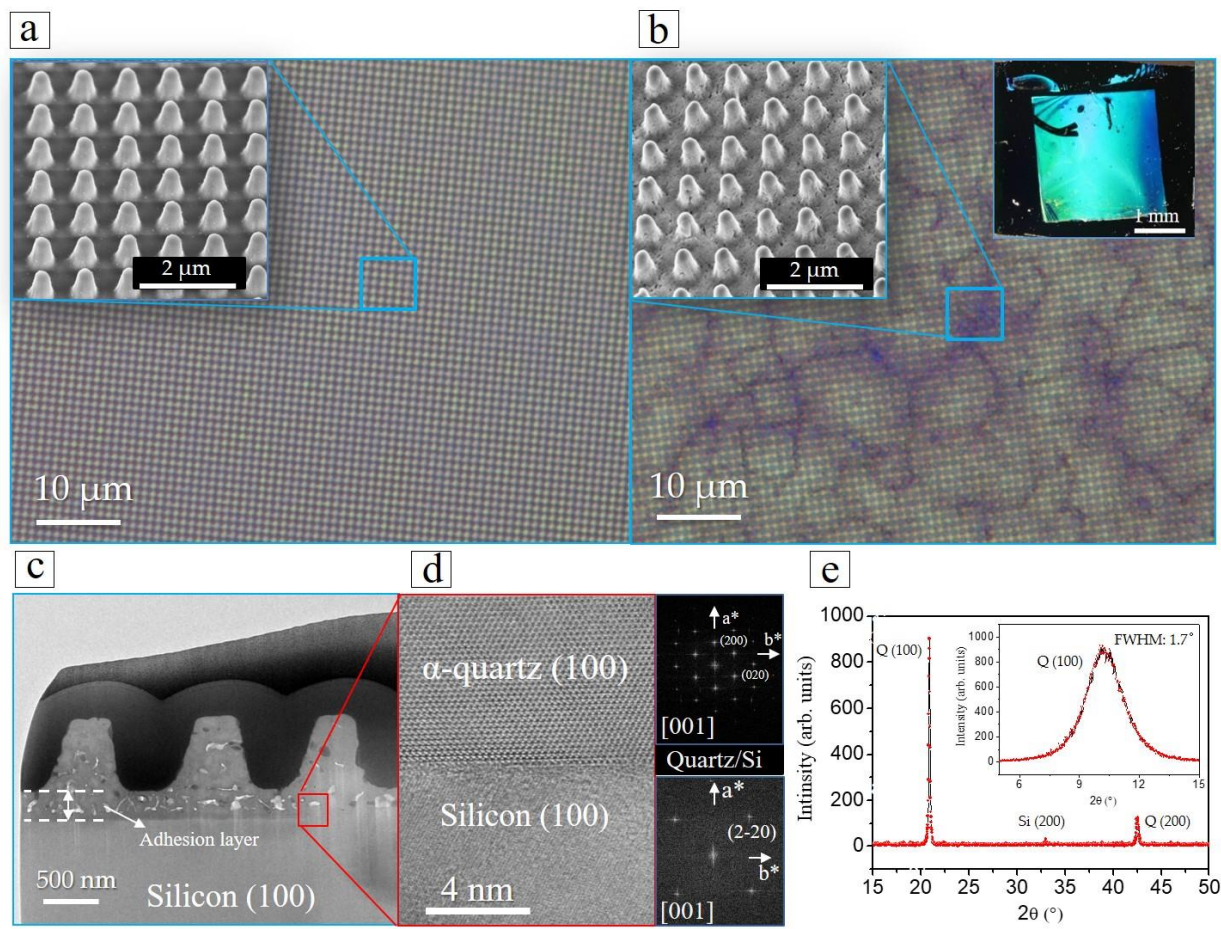

**Fig. 4. Crystallization of Sr-doped silica xerogel nanopattern.** (a) Optical image of Sr-doped silica xerogel nanopattern composed of 500 nm height columns performed by NIL lithographic process before crystallization process. The inset figure shows a higher magnification FEG-SEM image illustrating the morphology of the Sr-doped silica xerogel nanopattern. (b) Optical image the Sr-doped silica xerogel nanopattern sample after crystallization. Notice that is possible to observe the formation of typical quartz grain boundaries. The inset images shows a higher magnification FEG-SEM image of the quartz nanopattern (left side) and an optical images that exhibit the light diffraction after interaction with the quartz nanocolumn patter (right side). (c) Low magnification high angle annular dark field (HAADF) Z-contrast image of a quartz nanocolumn grown on the Si substrate assisted by the $Sr^{2+}$ catalyst at 1000 °C, 5 hours. (d) Atomic resolution Z-contrast image of a single (100)-oriented quartz nanocolumn viewed along the [100]-crystallographic direction. Inset figures show the corresponding FFT of both the quartz film and the silicon substrate. (d) θ-2θ XRD pattern with a perfect texture of the (100) α-quartz crystallographic phase after lithographic process. The inset shows a rocking curve showing a mosaicity value of quartz of 1.7° (e)

To evaluate the piezoelectricity of the nanoimprinted quartz columns we employed PFM. The obtained $d_{33}$ value was of the same order as the dense quartz films before lithographic process and the bulk material (see Fig. 5). The PFM amplitude image is represented in Figure 5a and the inset shows the topographic AFM image of crystallized nanocolumns. Notice that the areas surrounding the nanocolumns show a slight change in the PFM amplitude, the signal remains constant both at the top of the columnar structures at the quartz film surrounding the base of these pillars. The change of the PFM amplitude signal in the perimeter of the nanocolumns is attributed to a topographic crosstalk artifact which is well known and reported by the community[18]. We were able to corroborate the electromechanical behavior of our films by performing point-out spectroscopy measurements, see Fig. 5b. The electromechanical behavior of the structures was studied using frequency-sweeps to display the PFM contact resonant circuit. The electromechanical behavior is studied outside and inside of the nanocolumns by placing the AFM tip in each respective position. The data shows an increase of resonant amplitude with an increase of the applied AC bias, in a similar way as depicted in Fig. 2c, confirming that the nanostructuration has not been detrimental to the electromechanical properties.

Likewise, this nanostructuration methodology of epitaxial quartz thin films on silicon by NIL lithography is general for several kinds of patterns including lines as those shown in Fig. S8.

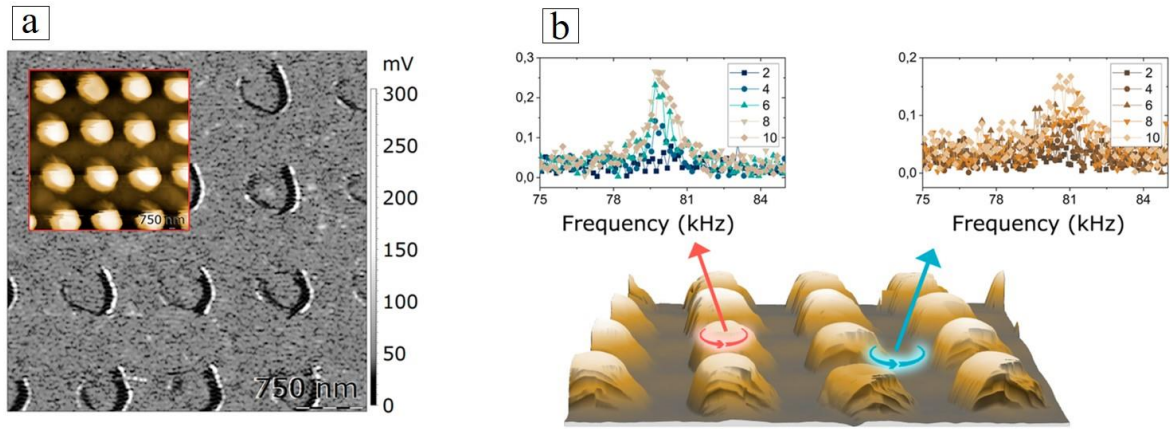

**Fig 5. Piezoelectric response of epitaxial nanostructured quartz films using NIL lithographic process. (a)** PFM amplitude and topography (inset) recorded simultaneously while applying a tip-substrate AC voltage of 10 V, showing area similar tip vibration level at the background film and top nanostructures. Point-out spectroscopy measurements recorded on top of the structures and bottom film, for different applied AC bias. **(b)** The data shows an increase of the PFM resonant frequency amplitude with an increase of the applied AC bias, confirming our expectation that piezoelectric functionality is preserved.

### 2.3. Self-assembled $SrCO_3$ nanoparticles as nanomasks for lithographic patterning of dense epitaxial quartz thin films on silicon.

Films with a $Sr/SiO_2$ molar ratio of 0.05 exhibit an outcropping of $SrCO_3$ nanoparticles at the surface, driven by a chemical reaction between SrO, $CO_2$ and $H_2O$, have been shown elsewhere[19]. These observations revealed the assembly of sintered $SrCO_3$ nanoparticles during the annealing treatment at 1000°C, whereas now, as illustrated in Fig. 6, we exploit these $SrCO_3$ nanoparticles as nanomasks to produce an array of quartz nanopillars from dense films. Indeed, solid $SrCO_3$ nanoparticles are extremely stabile under the reacting ion etching conditions. By this simple approach, illustrated in Fig. 1 and Fig. 6, one can produce arrays of quartz vertical nanopillars having diameters down to 60 nm and a maximum height of 400 nm, depending on the original quartz film thickness. This type of behavior has been reported for $CaF_2$ nanoparticles formed after a chemical reaction with the plasma etching[20]. In this case, the chemical transformation of the $Ca_xTi_{(1-x)}O_{(2-x)}$ present within the amorphous silica layer into the homogeneous dispersions of $CaF_2$ nanoparticles, was used as a particulate nanostencil system to produce an array of silicon nanopillars[20]. In our case, the $SrCO_3$ nanoparticles are formed during quartz crystallization and remained extremely stable during the RIE etching process, acting as an efficient nanomask that protects quartz from the plasma etching. This feature can be observed in Figure 6 that shows $SrCO_3$ nanoparticles before and after RIE process. Figure 6a and S9 show the typical morphology and size of the $SrCO_3$ nanoparticles on top of the epitaxial quartz thin film.

The efficacy of the process was investigated by electron microscopy and electron diffraction characterization was used to assess the crystalline structure of the quartz nanopillars (see figure 6). The etching has been applied in 100W RF and 200W LF of an inductively coupled plasma reactive ion etching (ICP-RIE) reactor using $CHF_3/O_2$ gas mixture (see more details in experimental section). The electron diffraction pattern of a single quartz nanopillar presented in Fig. 6b reveals perfect quartz crystallinity similar to that of the initial dense quartz film.

The morphology of the motifs is conical rather than needle-like as a result of an isotropic etching of $CHF_3/O_2$ flux. With the aim of producing networks of needle-like quartz nanostructures, we used ionized gases and gas mixtures such as Ar, CHF3, SF6 in order to control the anisotropy of the etching[21]. Unfortunately, under this etching conditions, quartz thin film and $SrCO_3$ nanoparticles were totally destroyed after 4 min (see figure SI 10).

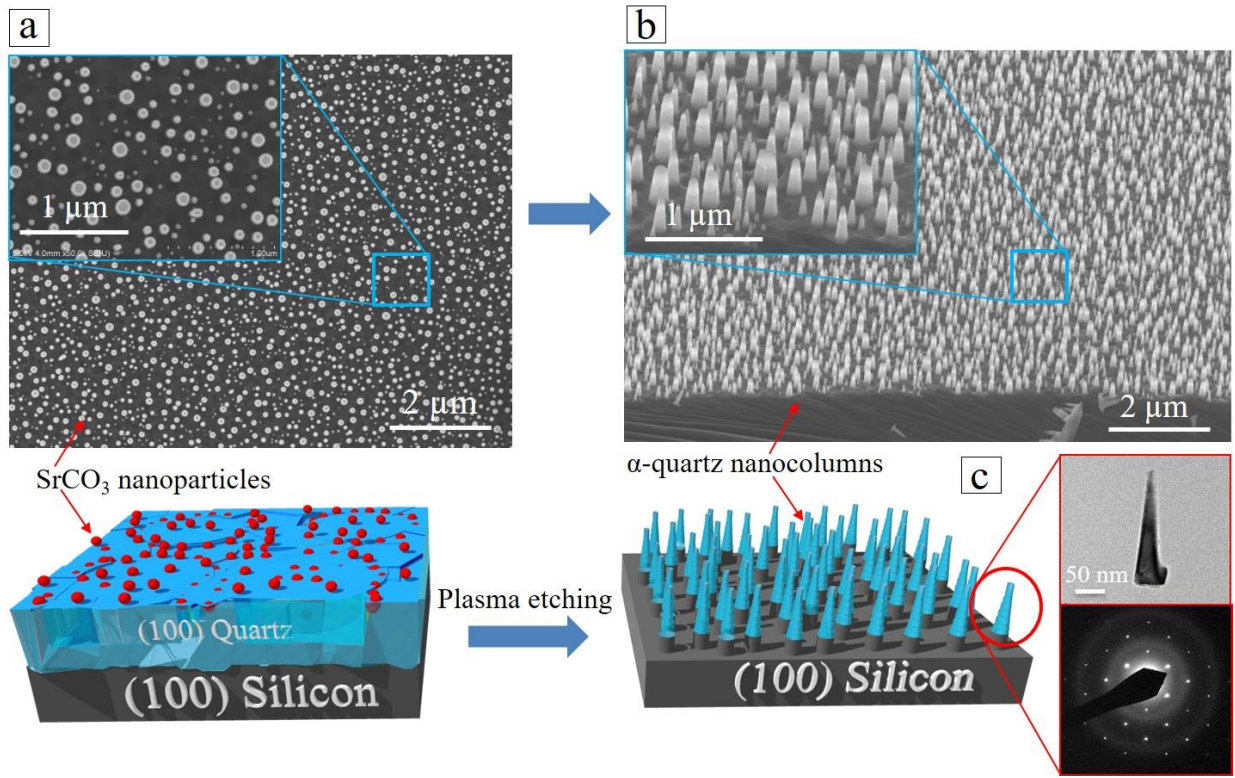

**Fig. 6. SrCO3 nanoparticles as nanomasks to produce an array of quartz nanopillars.** (a) FEG-SEM image illustrating the morphology of sintered SrCO$_3$ nanoparticles at 1000°C. The inset image shows a small zone of picture a with a higher magnification. (b) FEG-SEM image illustrating the morphology of quartz single crystal conical-like nanopillars after RIE etching of a dense film. The inset image shows a small zone of picture **b** with a higher magnification. General schematic of RIE etching to produce first nano lithographic patterns on epitaxial quartz thin films. (c) Electron diffraction pattern of a conical-like nanopillar in c shows a perfect crystallinity after the nanomask lithographic process.

3. **Conclusion**

The combination of top-down and bottom-up methodologies enabled the nanostructuration of piezoelectric quartz films, epitaxially grown on (100)-silicon substrates. We have used scalable lithographic methodologies that do not require masks to generate highly ordered 1D quartz patterns consisting of vertical quartz nanocolumns with diameters and heights ranging from 50 nm to 800 nm and 200 nm to 2 µm, respectively. The nanostructuration engineering of epitaxial quartz films on silicon presented here is general for several kind of patterns, being produced using exclusively low cost lithographic methodologies. LIL lithographic process allowed preparing quartz nanocolumns with diameters between 400 nm and 800 nm and heights in the range of 200 nm to 1000 nm, thanks to a novel multilayer film process that consists in the sequential deposition and consolidation of several gel layers. With this combination of methodologies, epitaxial 1D-quartz nanostructures maintain the crystallinity and epitaxial orientation of (100) α-quartz∥(100) Si. On the other hand, we have established the conditions to shown that NIL lithography combined with sol gel process is a versatile method to replicate several dimensions of Sr doped silica pillars depending on the characteristics of PDMS mould. Specifically, a withdrawal speed of 300 mm min$^{-1}$ at 25°C and 45% of humidity and the deposition of adhesion layer are required to obtain a perfect nano-imprinted crystalline continuous quartz pattern. Thus, the interplay between temperature, humidity, dip-coating conditions, and epitaxial growth plays a key role for the fabrication epitaxial quartz nanopillars on silicon substrates by NIL lithography. Finally, the controlled outcropping of $SrCO_3$ nanoparticles on top of epitaxial quartz thin films allows producing quartz vertical nanopillars having diameters down to 60 nm and a maximum height of 400 nm under the reacting ion etching conditions. In all cases, the nanostructuring of quartz films by no means detrimental to the piezoelectric properties of the film, which are preserved. We used two techniques,. DPFM and PFM, to quantify the piezoelectric coefficient $d_{33}$ of nanostructured and dense quartz films using in all cases a reference based in a commercial ferroelectric Periodically Poled Lithium Niobate (PPLN) sample.

This work demonstrates the integration on silicon of lithographic patterning of epitaxial quartz thin films, which precedes the production of microelectromechanical systems. The control at the nanoscale over the shape, micro- and nano-patterning of quartz thin films opens up the opportunity to fabricate new high frequency resonators and highly sensitive sensors relevant in different fields of application.

4. Experimental Section

**4.1. Synthesis**

Solution preparation: All the chemicals were from Sigma-Aldrich and without any further purification. In a typical process, we first prepared Solution A by adding 0.7 g Brij-58 into 23.26 g absolute ethanol, then 1.5 g HCl (37%), 4.22 g tetraethyl orthosilicate (TEOS) and stirring the solution for at least 4 h and not more than 18 h. After that, an aqueous solution of $Sr^{2+}$ was prepared with $SrCl_2 \cdot 6H_2O$ (Solution B). The solution used to prepare Sr-doped mesoporous silica films by dip-coating (Solution C) was obtained by adding 275 µL Solution B into 10 mL of as-prepared Solution A and stirring it for 10 min. The films were always obtained not later than 40 min after preparing solution C, as $Sr^{2+}$ is not stable in the latter. To obtain perfect and continuous crystallize quartz layers, we set the molarity of Solution B to 1M, which corresponds to a $Sr/SiO_2$ molar ratio of 0.05 resulting in a final molar composition of Solution C of TEOS:Brij-58:HCl:EtOH:$SrCl_2$=1:0.3:0.7:25:0.05.

Gel films by dip-coating: layer gel films on Si (100) substrates were prepared with a ND-DC300 dip-coater (Nadetech Innovations) equipped with an EBC10 Miniclima Device to control the surrounding temperature and relative humidity. During the dip-coating, we fixed the ambient temperature and relative humidity as 25ºC and 40% and the thickness of film was controlled by the withdrawal rate. In this study, all the films were made at withdrawal rate of 300 mm/min to ensure the perfect crystallization and nanoimprint process. After dip-coating, as-prepared gel films were consolidated with a thermal treatment of 5 min at 450 ºC under air atmosphere. In order to reach thicker films, and therefore taller nanocolumns, multi-layer gel films were obtained by repeating the required number of times the process of mono-layer preparation on the same substrate.

*Crystallization:* As-prepared gel films were introduced into a furnace already at 1000ºC in air atmosphere and held at this temperature for 300 min. The crystallized films were recovered after natural cooling of the furnace to room temperature.

### 4.2. Structural Characterization and Piezoelectric Measurements

X-Ray Diffraction (XRD): The crystalline textures and rocking curve measurements of films were performed on a Bruker D8 diffractometer (3 s acquisition every 0.02º in Bragg-Brentano geometry, with a radiation wavelength of 0.154056 nm). Epitaxial relationship was analyzed through X-ray diffraction measurements by using a Bruker AXS GADDS equipped with a 2D X-ray detector. Optical Microscope: The optical images of films were obtained in an Olympus BX51M optical microscope equipped with a Nikon DS-Fi3 camera. Field Emission Gun Scanning Electron Microscopy (FEG-SEM): The microstructures of the films were investigated with a FEG-SEM model Su-70 Hitachi, equipped with an EDX detector X-max 50 mm$^2$ from Oxford instruments. Transmission Electron Microscopy (TEM): Cross-sectional studies of films were performed by using a FEI Titan3 operated at 80 kV and equipped with a superTwin® objective lens and a CETCOR Cs-objective corrector from CEOS Company. Electron diffraction studies were performed in a JEOL 1210 operated at 120 kV. Atomic Force Microscopy (AFM): The topography of nanostructured and dense quartz films was studied by AFM in a Park Systems NX-Scanning Probe Microscopy (SPM) unit. Piezoelectric characterization through the direct piezoelectric effect was made by Direct Piezoelectric Force Microscopy[15] in an Agilent 5500LS instrument using a low leakage amplifier (Analog Devices ADA4530) with Platinum solid tips (Rockymountain Nanotechnology RMN-25 PtIr200H). PFM measurements were performed in an

Agilent 5500LS using a long-tip shank length tip[22] to diminish electrostatic interaction (RMN 25PtIr300b) while working in the resonant frequency (∼ 80 kHz). A Periodically Poled Lithium Niobate from Bruker AFM was used as a reference testing platform.

### 4.3. Quartz thin films nanostructuration

**Optical lithography**. First set of samples has been fabricated by top-down approach using conventional optical photolithography following by anisotropic plasma etching. First, linear micrometer scale patterns have been insolated in 1.1 µm thick photoresist layer (ECI from MicroChemicals) using a conventional mask aligner (SUSS Mask Aligner MA 6). The patterns where transferred on epitaxial quartz thin films (see figure S1) by performing plasma etching using fluoroform chemistry, low pressure (5 mTorr), and 100 W bias power in pure capacitive coupling plasma (CCP). Finally, the remaining ECI resist was stripped in acetone and rinsed in isopropanol.

**Laser interferential lithography (LIL)**

In order to produce an epitaxial quartz nanocolumn pattern from dense films, we used a positive photoresist, AZ MIR 701, which was exposed using the interferential lithography technique to obtain a network of dots after the using a developer, AZ726. This procedure allows to rapidly obtain periodic design over a large surface (∼cm$^2$) without the need of a lithographic mask[7]. For the quartz pattern in Figure 2, a 405 nm wavelength laser with a divergent beam was reflected by two mirrors shifted with an angle of 10° which resulted in an interferometric pattern with a pitch of 1 µm. To obtain the dot pattern two exposures were needed, a first exposure created periodic lines and a second exposure shifted by 90° with respect to the first exposure generated perpendicular periodic lines. The result of these two exposures, after development, generates the dots. Finally, the samples were anisotropically etched by inductively coupled plasma reactive ion etching (ICP-RIE) (model corial 200 IL) using $CHF_3/O_2$ gas mixture. RIE conditions for etching of the sample and then produce a periodic pattern of quartz pillars of 1 µm depth were the following: power:120W RF, 400 W LF, gas: $CHF_3$ 100 sccm-$O_2$ 20 sccm (standard cubic centimeter per minutes), pressure: 10 mTorr and time: 10 min. ICP-RIE produces a dry and directional etching induced by a mixture of $CHF_3$ and $O_2$ plasma.

**Soft nano-imprint lithography (NIL) Preparation**

*Moulds preparation:* Si masters were elaborated with different structures and heights using LIL lithography. PDMS (polydimethylsiloxane) reactants (90 w% RTV141A; 10 w% RTV141B from BLUESIL) were transferred onto the master and dried at 70 °C for 1 h before unmoulding.

Then, a first silica layer seed was deposited at a constant relative humidity of 45% with controlled withdrawal speeds of 300 mm $min^{-1}$ in order to adjust the final thickness to 200 nm, and was consolidated at 450°C for 10 min. Importantly, this layer has two different functionalities: (i) as a seed layer to produce a continuous and homogeneous epitaxial quartz thin film on silicon and (ii) as an adhesion layer to faultlessly replicate the columnar shape from the PDMS mould. Then, a new layer of the same solution was deposited under the same conditions for printing. Surfactant Brij-58 included in the final sol-gel Solution C did not change the wetting properties of the sol.

After the last dip-coating, the substrates were quickly introduced during 1 min into a custom-designed chamber under a controlled temperature of 25 °C using and a constant humidity of 45%. Imprinting of sol–gel films with a PDMS mould involves the following steps. First, moulds were degassed under vacuum (10 mbar) for 20 min before direct application on the as-prepared xerogel films kept in a controlled environment, without additional pressure. After 1 min, the samples were transferred to a 70 °C stove for 2 min and then to a 120 °C stove for 2 min to consolidate the xerogel films before peeling off the PDMS mould. Next, the sol–gel replicas were annealed at

450 °C for 10 min for consolidation. Finally, sample was crystallized at 1000ºC for 5h in air atmosphere.

**Nanomasks lithography**

Quartz samples covered by $SrCO_3$ nanomasks were anisotropically etched by inductively coupled plasma reactive ion etching (ICP-RIE) (model corial 200L) using $CHF_3/O_2$ gas mixture. The RIE conditions to engrave the sample and then produce quartz nanopillars pattern of 400 nm depth were the following: power:120W RF, 400 W LF, gas: $CHF_3$ 100 sccm-$O_2$ 20 sccm , flux:, pressure: 10 mTorr and time: 10 min.

Isotropic etching conditions used in samples of Fig. S10 were the following: power: 100W RF, 200 W LF, gas: $CHF_3$ 80 sccm-$O_2$ 10 sccm , SF6 20 sccm, pressure: 10 mtorr and time: 4 min ( for a 300 nm thick quartz layer).

Notice that if requested the $SrCO_3$ nanoparticles can be dissolved by dipping the sample into a nitric acid solution (3 M) for 2 hours after quartz crystallization[19].

**Acknowledgements**

This project has received funding from the European Research Council (ERC) under the European Union's Horizon 2020 research and innovation programme (No.803004); the French Agence Nationale pour la Recherche (ANR), project Q-NOSS ANR ANR-16-CE09-0006-01; the Spanish Ministry of Science Innovation and Universities in co-funding with European Social funds through the Severo Ochoa Program for Centers of Excellence in R&D (SEV-2015-0496) and the Ramón y Cajal program (J.G., RyC-2012-11709); the Generalitat de Catalunya (2017SGR00765). Q.Z. was financially supported by China Scholarship Council (CSC) with No.201506060170. Q.Z.'s work was done as a part of the Ph.D program in Materials Science at Universitat Autònoma de Barcelona. The authors thank the "Laboratorio de Microscopías Avanzadas-Instituto de Nanociencia de Aragón" for offering their expertise in the preparation of TEM cross-sections. A. Crespi from XRD diffraction service is acknowledged for pole figure measurements. FEGSEM instrumentation was facilitated by the Institut des Matériaux de Paris Centre (IMPC FR2482) and was funded by Sorbonne Université, CNRS and by the C'Nano projects of the Région Ile-de-France. We thank David Montero for performing the FEGSEM images.

# Nanostructure engineering of epitaxial piezoelectric α-quartz thin films on silicon

*Q. Zhang[1, 2 γ], D. Sánchez[1 γ], R. Desgarceaux[1 γ], A. Gomez[2], P. Escofet-Majoral[1], J. Oró-soler[2], J. Gazquez[2], G. Larrieu[3], B. Charlot[1], M. Gich[2], A. Carretero-Genevrier[1]\**

1. Rudy Desgarceaux, Pau Escofet-Majoral, Dr. Benoit Charlot, Dr. Adrian Carretero-Genevrier
Institut d'Electronique et des Systemes (IES), CNRS, Universite Montpellier 2 860 Rue de Saint Priest 34095 Montpellier, France
2. Qianzhe Zhang, Andres Gomez, Dr. Jaume Gazquez, Dr. Marti Gich
Institut de Ciència de Materials de Barcelona ICMAB, Consejo Superior de Investigaciones Científicas CSIC, Campus UAB 08193 Bellaterra, Catalonia, Spain
3. Guilliem Larrieu
 LAAS, Université de Toulouse, CNRS, INP, Toulouse, France

γ These authors contributed equally to this work
E-mail: carretero@ies.univ-montp2.fr,

Keywords: Quartz, silicon, epitaxial growth, thin films, piezoelectricity, nanostructuration

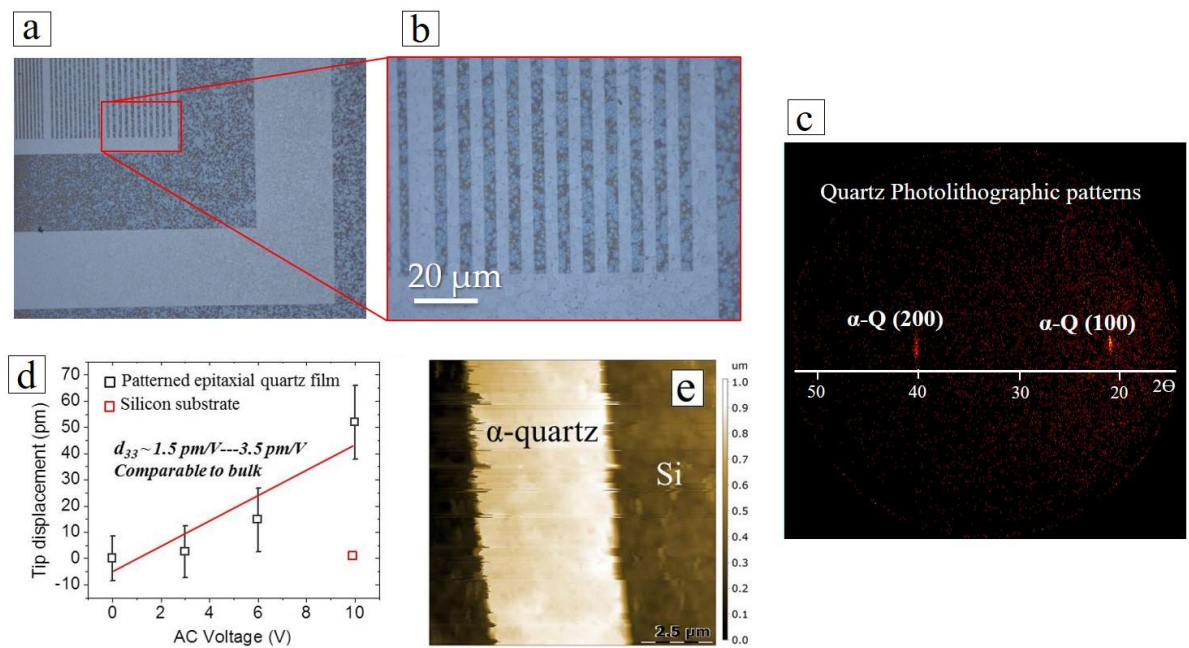

**Fig. S1. Photo-lithographic patterning of piezoelectric quartz thin films on silicon substrates.**
(**a** and **b**) Lithographic patterning of epitaxial textured quartz thin films using anisotropic plasma etching. (**c**) 2D diffraction pattern lithographed quartz sample. (**d** and **e**) PFM measurements of patterned epitaxial quartz films.

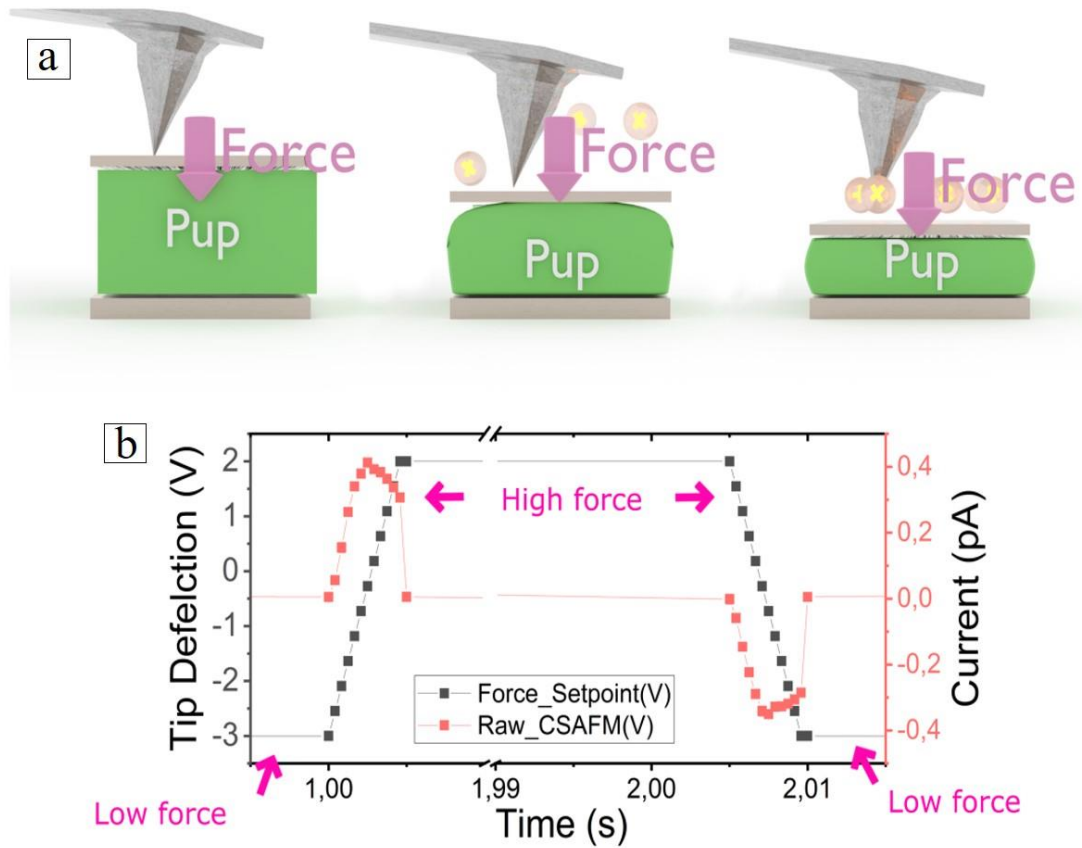

**Fig. S2. DPFM measurements on quartz dense films. (a)** Force and Piezoelectric current vs Time (s) for different applied forces on a 800 nm thick quartz film before lithographic process. Notice that the force profile starts with a constant value, while at 1s the force is increased to a value of 380 µN in a 5 ms time step. Following this step, a constant force is re-settled for an additional 1s, while at 2s, an unloading ramp is performed, reducing the applied force. While force is varied, the current channel is recorded simultaneously; current is depicted in red squares. Scheme of the spectroscopy experiments carried out in which the AFM tip applies a suitable force, within a given time, collecting the charges generated by the direct piezoelectric effect. **(b)** Notice that a constant force builds up a constant charge, hence the recorded current remains zero. However, when the force is varied, through a loading or unloading event, there is an increase or decrease of the charge build up, whereas a constant current can be seen at a constant force rate applied. Curves performed in α-quartz dense film showing its piezoelectric response. The graphs were obtained by averaging 4x4 matrix volume spectroscopy experiments in an area of 10 microns, in order to depict the homogeneity of the sample. From these measurements we obtained a $d_{33}$ of 4 ± 2 pC/N in agreement with the piezoelectric coefficient of α-quartz. Notice that DPFM methodology cannot be applied to nanostructured quartz films because the applied force breaks

quartz nanocolumns, making impossible this kind of measurement. However, we recently combined PFM and DPFM methodology to quantify the piezoelectric coefficient $d_{33}$ at the nanoscale in $BiFeO_3$ ferroelectric epitaxial oxide thin films[1], therefore validating this approach.

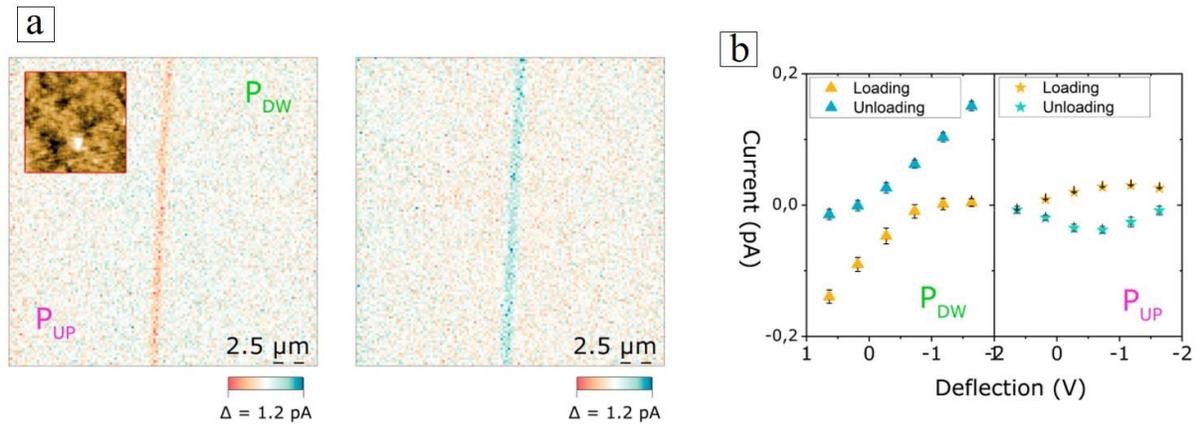

**Fig. S3. DPFM on PPLN tests sample.** **(a)** Ferroelectric Periodically Poled Lithium Niobate (PPLN) sample was used as a reference. For obtaining the location of ferroelectric domains, DPFM-Si (Signal Input) and DPFM-So (Signal Output) scans were recorded, revealing the expected antiparallel domain configuration for a ferroelectric known sample[2]. **(b)** Results of the spectroscopy experiments, obtained under similar conditions to those used for quartz thin films, in four different locations on the PPLN tests sample indicated in the an upwards polarization ($P_{UP}$) and downwards polarization ($P_{DW}$).

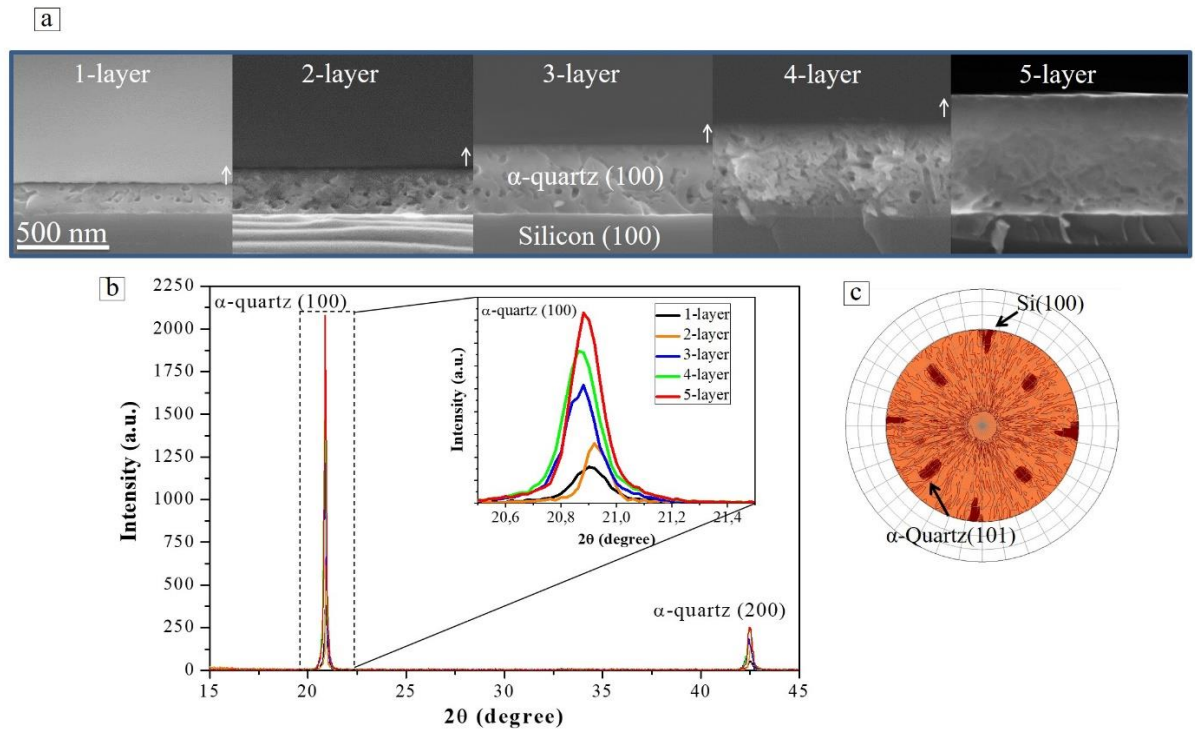

**Fig. S4. FEG-SEM images of the cross section of the films with different number of layers. (a)** Note that the thickness of the different multilayers films corresponds to the final height of the quartz columns plotted in figure 2e of the main manuscript. **(b)** XRD θ - 2θ scan showing the linear increase of the intensity of the α-quartz (100) reflections as the number of layers increases. **(c)** Pole figure for 5-layer film to show that the epitaxial growth is maintained.

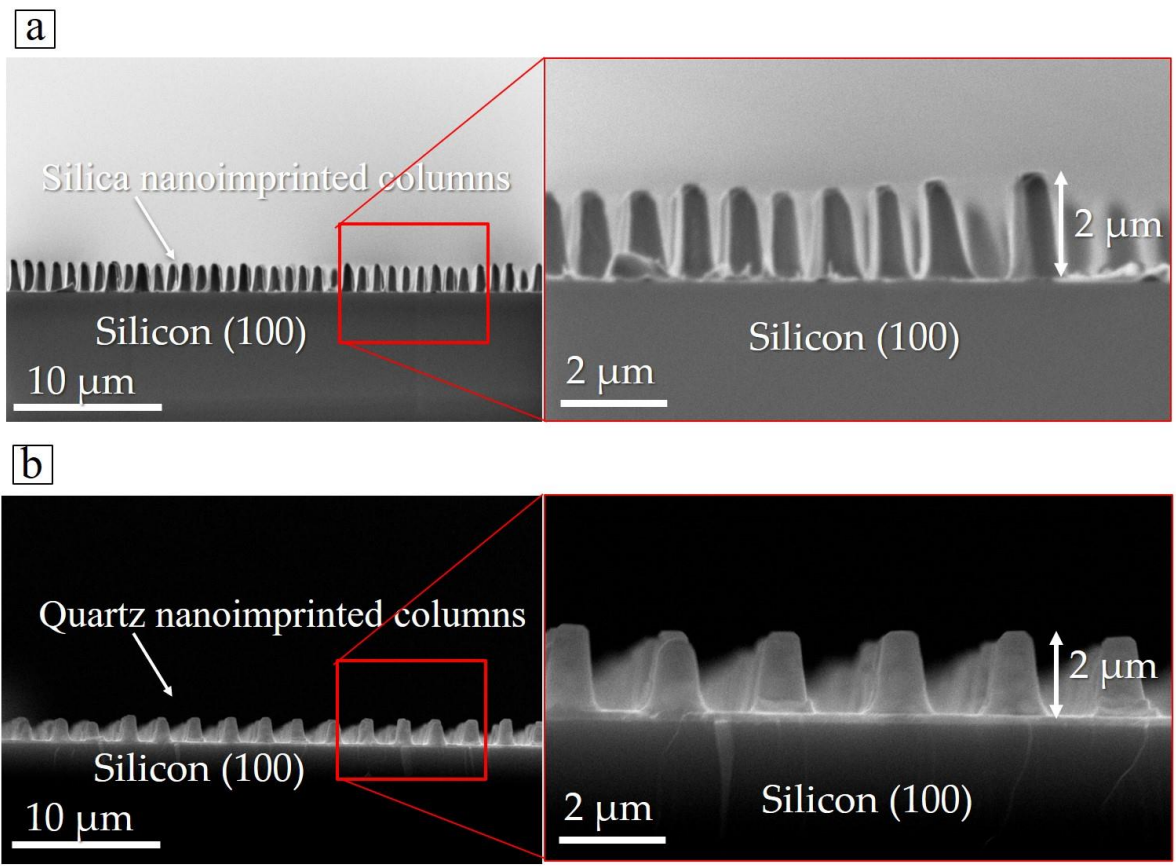

**Figure S5.** (**a**) SEM cross-sectional images of 2 μm height Sr doped silica nanopillars before crystallization. (**b**) SEM image of 2 μm height α-quartz nanopillars grown at 1000 °C for 5 hours in air atmosphere.

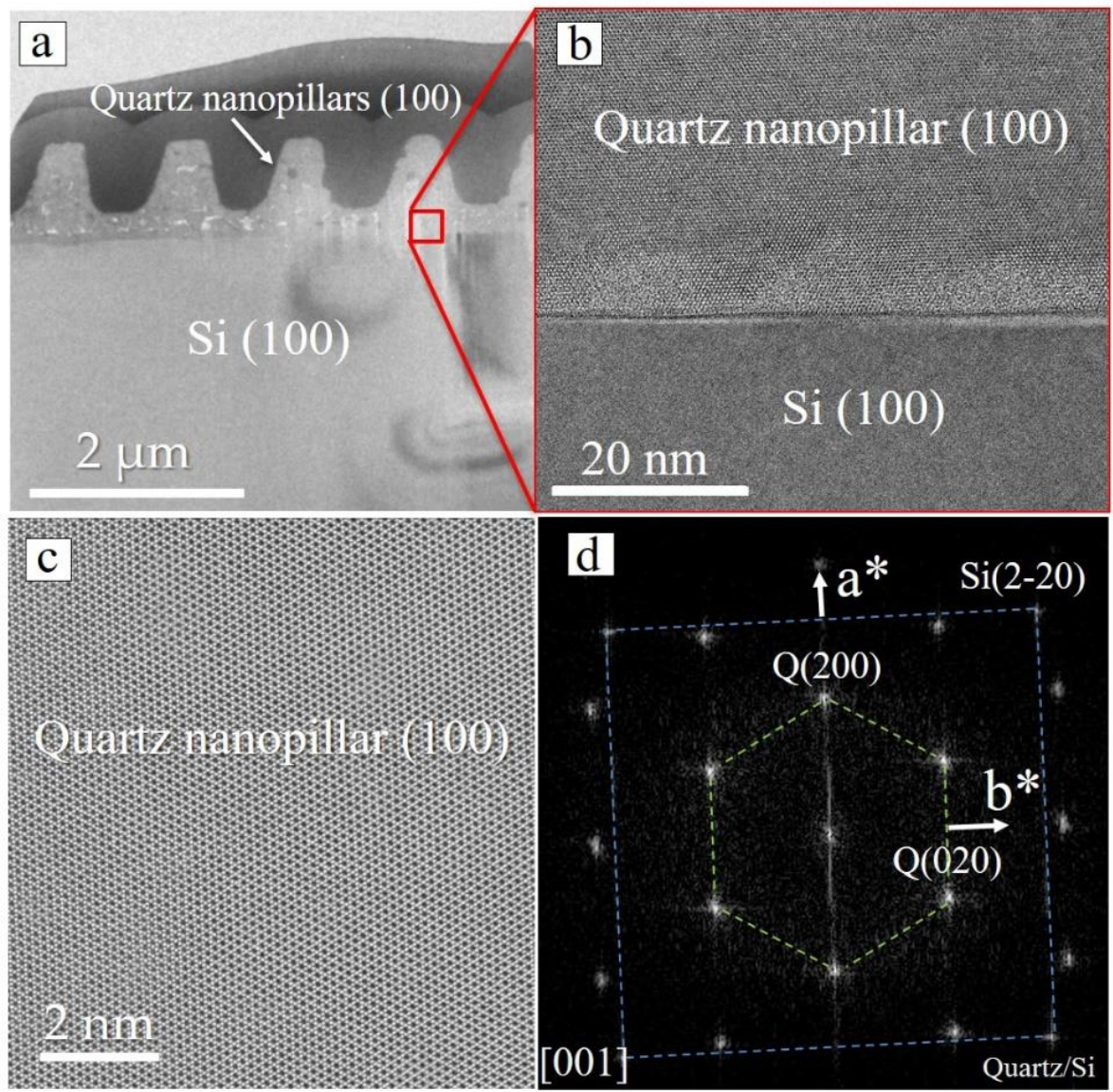

**Figure S6. Advanced Structural Characterization of nanostructured quartz films.** (**a**) Low magnification high angle annular dark field (HAADF) Z-contrast image of epitaxial quartz nanocolumns grown on the Si substrate assisted by the $Sr^{2+}$ catalyst at 1000 °C, 5 hours. (**b**) Atomic resolution Z-contrast image of Quartz/silicon interface viewed along the [100]-crystallographic direction. (**c**) Atomic resolution Z-contrast image of a single (100)-oriented quartz nanocolumn viewed along the [100]-crystallographic direction. (**d**) FFT of both the quartz film (green dashed line) and the silicon substrate (blue dashed line).

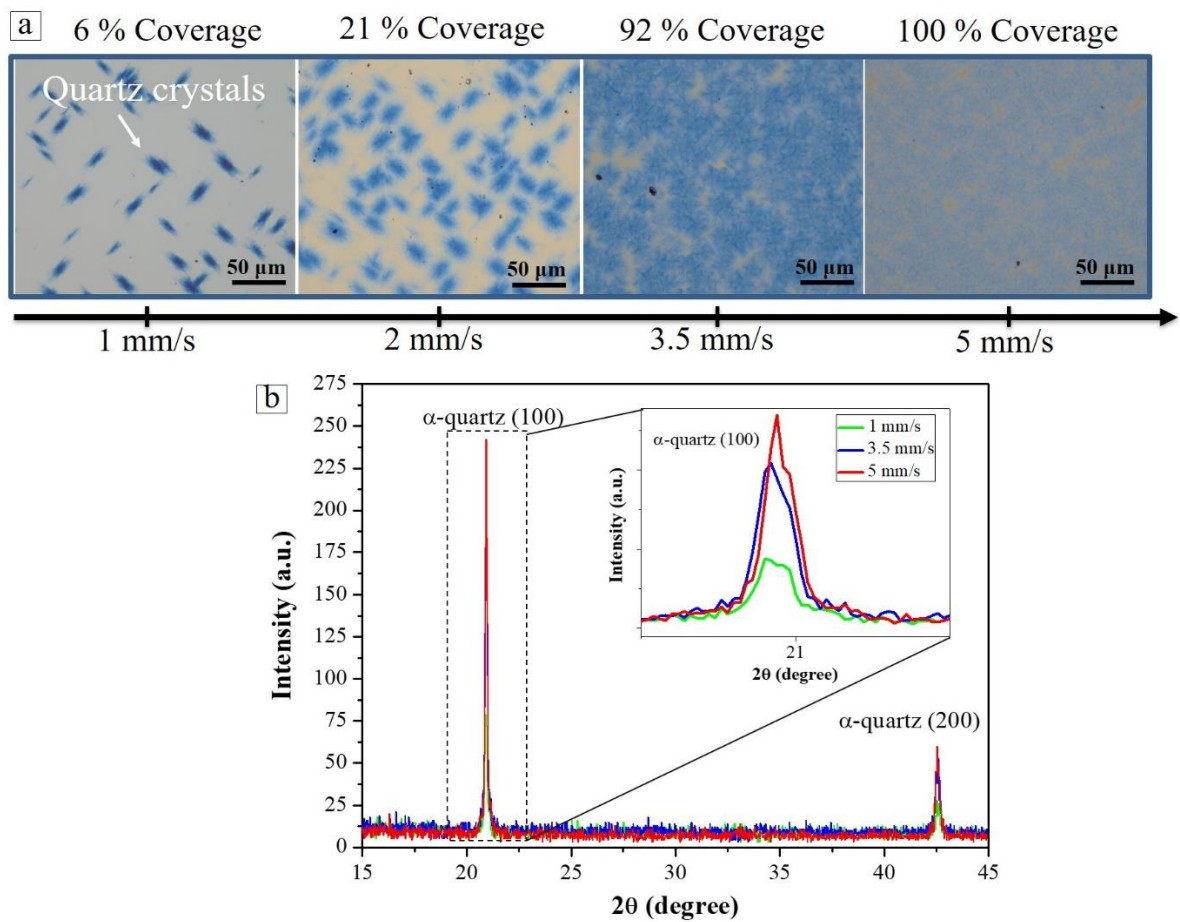

**Figure S7. (a)** Optical images of films prepared with different withdrawal speeds (indicated in the axis below the images); **(b)** The areal coverages of α-quartz crystals is indicated next to the images. The corresponding XRD θ-2θ measurements are shown in **(c)**. Those films were prepared in a relative humidity of 40% at 25°C using Brij-58 as surfactant. The annealing treatment was at 1000°C for 300 minutes in air atmosphere.

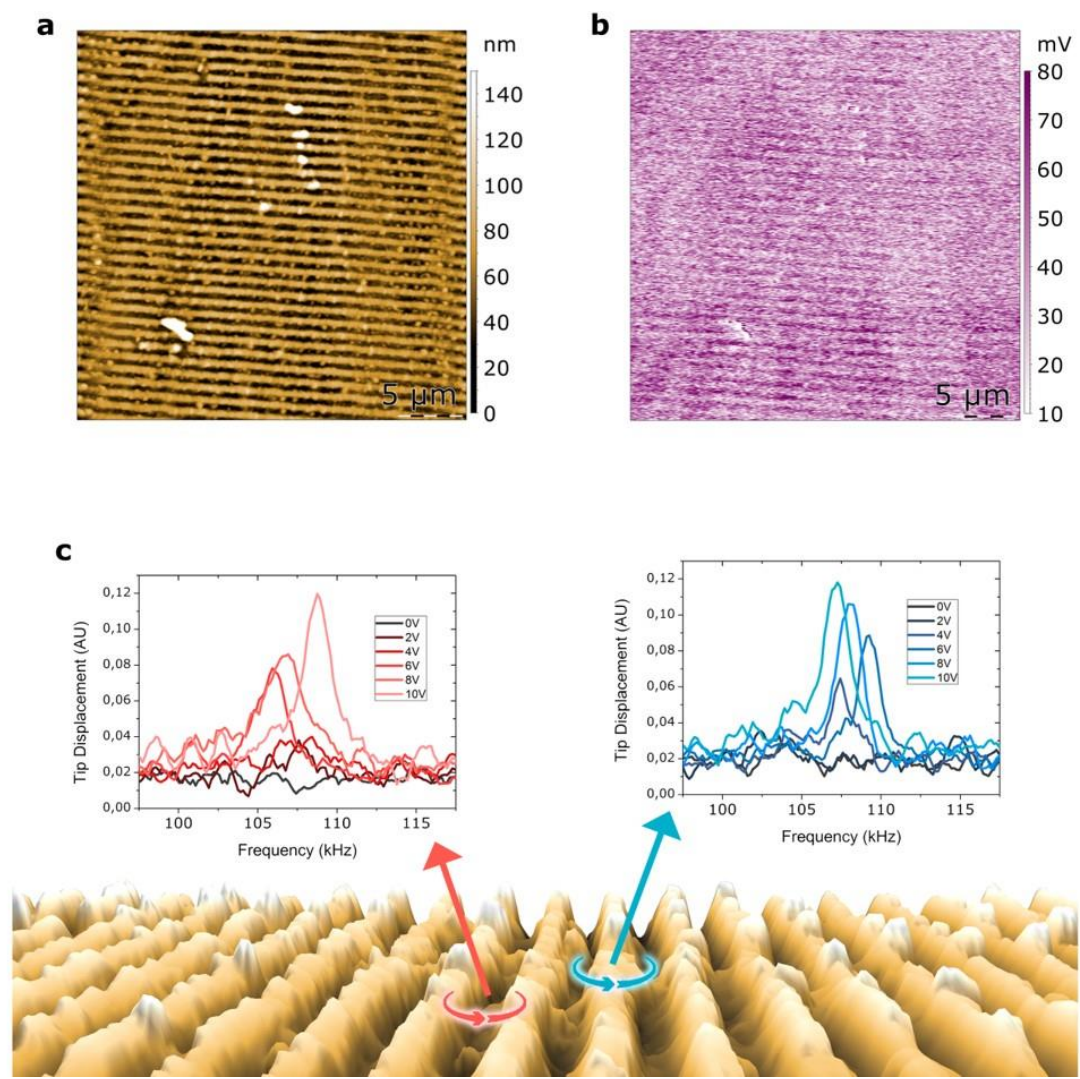

**Fig S8**. **Piezoelectric response of epitaxial quartz lines using NIL lithographic process.** (**a**) AFM image of the topography and PFM amplitude recorded simultaneously under a tip-substrate AC voltage of 10 V, showing the relationship between applied voltage and tip deflection on epitaxial quartz nanostructure and surrender area. Notice that the PFM response preserves the features of the topographic image, namely the quartz crystals surrounding the nanoline and the top of the nanoline. (**b**) We found no significant differences between PFM response within and outside the nanostripes. Thus, the piezoelectric functionality of the material is completely preserved.

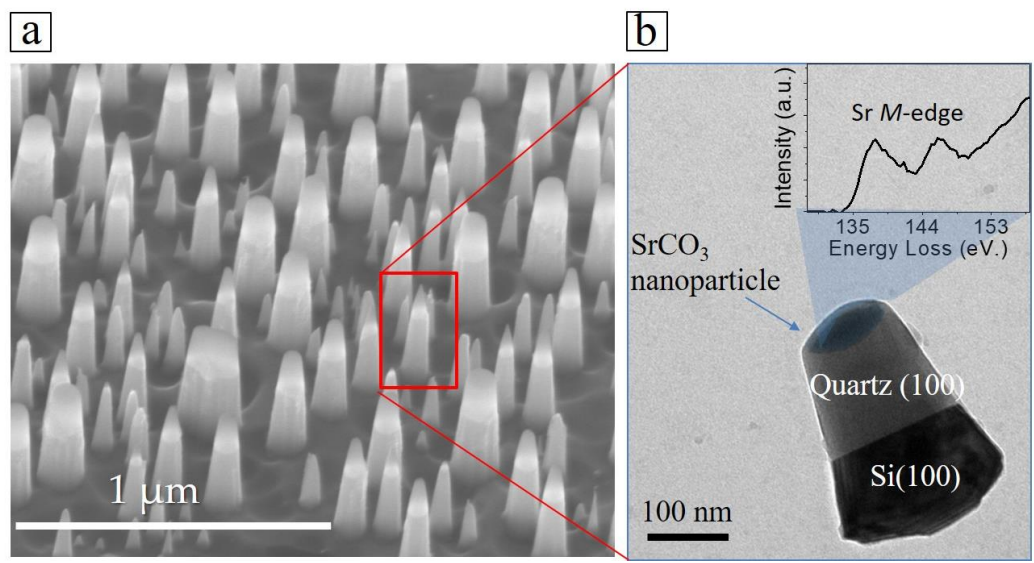

**Fig. S9. (a)** FEG-SEM image illustrating the morphology of nanocolumns obtained by nanomask lithography. **(b)** TEM image illustrating the morphology of quartz single crystal conical-like nanopillars after RIE etching of a dense film. The inset image shows the Sr M-edge EEL spectrum acquired from the $SrCO_3$ nanoparticle placed on top of the quartz nanocolumn.

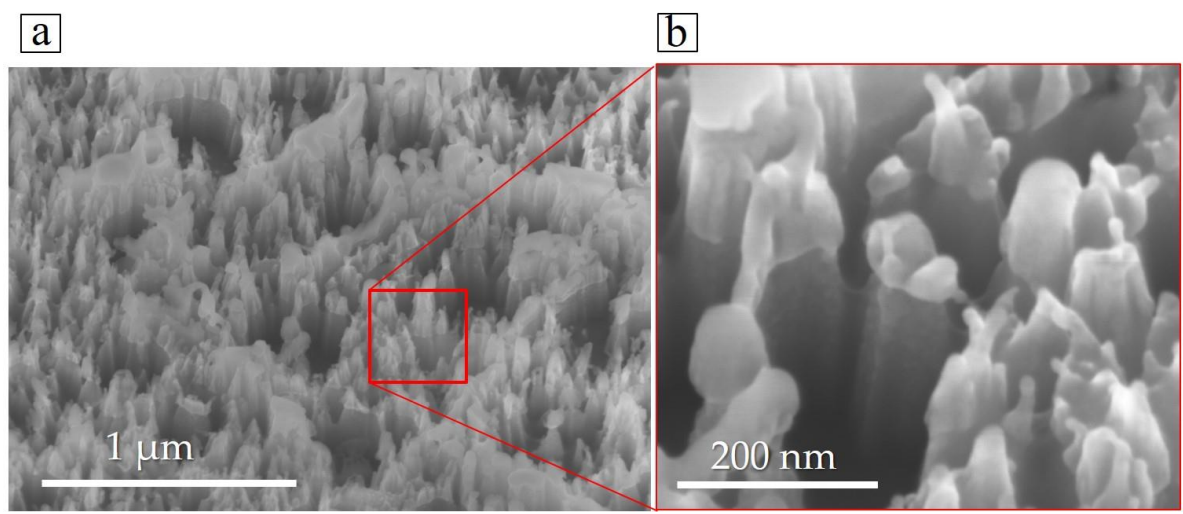

**Fig. S10. (a and b)** FEG-SEM image illustrating how quartz surface is destroyed after using Gas mixtures: CHF3, SF6. The failed objective of this experiment was to control the anisotropy of the etching.